\documentclass[10pt,fleqn,twocolumn]{revtex4}

\usepackage{tikz}
\usepackage[T1]{fontenc}
\usepackage{graphicx}
\usepackage{amssymb}
\usepackage{times}
\usepackage{color,amssymb,mathptm,times}
\usepackage{amsfonts,amssymb,latexsym,amsmath,amscd}

\newcommand{\bed}{\[}
\newcommand{\eed}{\]}
\newcommand{\beq}{\begin{equation}}
\newcommand{\eeq}{\end{equation}}
\newcommand{\beqa}{\begin{eqnarray}}
\newcommand{\eeqa}{\end{eqnarray}}
\newcommand{\ket} [1] {\vert #1 \rangle}
\newcommand{\bra} [1] {\langle #1 \vert}
\newcommand{\braket}[2]{\langle #1 | #2 \rangle}

\newcommand{\mean}[1]{\langle #1 \rangle}

\newcommand{\tr}{\mathop{\mathrm{tr}}}

\newcommand{\be}{\begin{eqnarray}}
\newcommand{\ee}{\end{eqnarray}}
\newcommand{\bea}{\begin{eqnarray}}
\newcommand{\eea}{\end{eqnarray}}
\newcommand{\bma}{\begin{subequations}}
\newcommand{\ema}{\end{subequations}}

    {
    \smallskip
    \refstepcounter{theorem}
    \noindent
    {\bf Example \Roman{section}.\arabic{theorem}} \ \ }
    {
    \smallskip}

    {
    \smallskip
    \refstepcounter{theorem}
    \noindent
    {\bf Remark \Roman{section}.\arabic{theorem}} \ \ }
    {\hspace*{\fill}
    \smallskip}

\newenvironment{definition}
    {
    \smallskip
    \refstepcounter{theorem}
    \noindent
    {\bf Definition \Roman{section}.\arabic{theorem}} \ \ }
    {\hspace*{\fill}{\ }
    \smallskip}

    {
    \smallskip
    \refstepcounter{theorem}
    \noindent
    {\bf Definition \Alph{section}.\arabic{theorem}} \ \ }
    {\hspace*{\fill}{\ }
    \smallskip}

    {
    \smallskip
    \refstepcounter{theorem}
    \noindent
    {\bf Scholium \Roman{section}.\arabic{theorem}} \it \ \ }
    {\hspace*{\fill}{\ }
    \smallskip}

    {
    \noindent
    {\bf Proof{#1}:  }
    }
    {\hspace*{\fill}{$\Box$}\smallskip}

    {
    \noindent
    {\bf Sketch{#1}:  }
    }
    {\hspace*{\fill}{$\Box$}\smallskip}

    {
    \noindent
    {\bf Reference:  }
    }
    {\hspace*{\fill}{$\odot$}\smallskip}

\newtheorem{theorem}{Theorem}[section]

\newtheorem{lemma}[theorem]{Lemma}

\usepackage{graphicx}
\usepackage[mathscr]{eucal}
\usepackage{amsmath}
\usepackage{amssymb}
\usepackage{epstopdf}
\usepackage{epsfig}
\usepackage{wrapfig}

\def\one{\ensuremath{\hbox{$\mathrm I$\kern-.6em$\mathrm 1$}}}
\def\tr{ \mbox{tr}}

\begin{document}

\title{Simulated annealing for tensor network states}
\author{S. Iblisdir}
\affiliation{Dept. Estructura i Constituents de la Mat\`eria, Universitat de Barcelona, 08028 Barcelona, Spain}

\date{\today}  


\begin{abstract}

Markov chains for probability distributions related to matrix product states and 1D Hamiltonians are introduced. With appropriate 'inverse temperature' schedules, these chains can be combined into a simulated annealing scheme for ground states of such Hamiltonians. Numerical experiments suggest that a linear, i.e. \emph{fast}, schedule is possible in non-trivial cases. A natural extension of these chains to 2D settings is next presented and tested. The obtained results compare well with Euclidean evolution. The proposed Markov chains are easy to implement and are inherently sign problem free (even for fermionic degrees of freedom). 

\end{abstract}

\maketitle

\textbf{\emph{Introduction.--}} Local quantum hamiltonians are central to our understanding of condensed matter systems. Our description of exotic phenomena, such as superconductivity or quantum magnetism, relies on finely tuned models for short-range, particle-particle interactions. Although simple in their expression and interpretation, such hamiltonians are generally difficult to study; most of them cannot be solved exactly or perturbatively, while entanglement makes mean field methods conceptually inadequate for them.  

In this respect, tensor network states have risen as very powerful instruments during the last two decades. The notion of matrix product states \cite{review-mps}, which underlies the celebrated density matrix renormalisation group method $(\textsf{DMRG})$ \cite{DMRG}, is now regarded as fundamental when dealing with gapped quantum spin chains. Extensions of this notion to two dimensional or to critical systems are reshaping our views on issues such as topological order \cite{peps-topo,mera-topo} or conformal invariance \cite{mera-cft}. There seems to be a deep reason for the adequacy of tensor network states at describing exotic phases; these exhibit finite but short-range entanglement. Area laws express this fact in precise statements \cite{area-law}.

The study of strongly correlated systems with tensor network states raises a fundamental issue, which we will refer to as the Optimal State Problem $(\textsf{DMRG})$. It can be informally formulated as follows. \emph{Given a particular class of tensor network states and a quantum hamiltonian, how to find the best approximation to the ground state of this Hamiltonian within that class?} The physical relevance of this problem is obvious; important frustrated or superconducting systems are modelled by hard-to-solve hamiltonians. But it has also aroused interest from a complexity theory perspective. In the case of matrix product states, some non-trivial instances of that problem are now known to belong to the $\textsf{P}$ complexity class \cite{gapped-h-in-P}, while others have been shown to be $\textsf{NP}$-hard \cite{Schuch:NP}. The Euclidean evolution ($\textsf{EE}$) or local optimisation method \cite{peps} are examples \emph{practical} of methods to tackle the optimal state problem and they have proven hugely successful in a wide variety of situations. But they are all deterministic or greedy in some sense. As a result, they sometimes get stuck when applied to Hamiltonians which energy landscape is plagued with local minima, as very simple examples show \cite{AAI}. The algorithms proposed in \cite{AAI,gapped-h-in-P} are conceptual breakthroughs in this respect; their validity is however limited to one-dimensional uniformly gapped hamiltonians, and their implementations seem very tedious in practice.  

The present paper is meant as a contribution to a systematic investigation of the optimal state problem. We here wish to depart from previously proposed search methods, and introduce a family of approximation schemes for quantum Hamiltonians, based on the Metropolis algorithm \footnote{Certainly, some Monte-Carlo schemes for tensor network states already exist \cite{V-MC-TPS}. But they were designed for a quite different purpose than ours. The focus in these early works was on efficient evaluation of observables, i.e. tensor contraction; the search of optimal states was carried out through appropriate estimation of \emph{gradients}. Here, we wish to take a different route altogether; standard evaluation of observables will do, but we will now aim at genuinely stochastic optimisation of states, where moves towards \emph{worse} configurations will be allowed.}. These schemes are flexible, efficient, and \emph{are relatively easy to implement}. We will first consider one-dimensional Hamiltonians and matrix product states; a \emph{simulated annealing} algorithm for the optimal state problem will be introduced, and its validity will be demonstrated in a concrete case study: a finite Ising chain. We will next move on to describe a natural adaptation of the algorithm to two dimensions, and an implementation for Ising models on a $7 \times 7$ square lattice, where the scheme will be shown to compare well with standard search methods. An important advantage of our Markov chains, when dealing with two-dimensional systems, is the possibility to include the error made when evaluating mean values of observables \cite{peps}. This feature results in \emph{inherently stable} algorithms yielding \emph{rigorous} upper bounds on ground state energies. We conclude with a discussion of some open questions and possible improvements and extensions; the appendices contain a brief discussion of convergence and some technical aspects related to the two-dimensional computations.

\begin{figure}[h]
\begin{center}
\includegraphics[width=70mm,height=17mm]{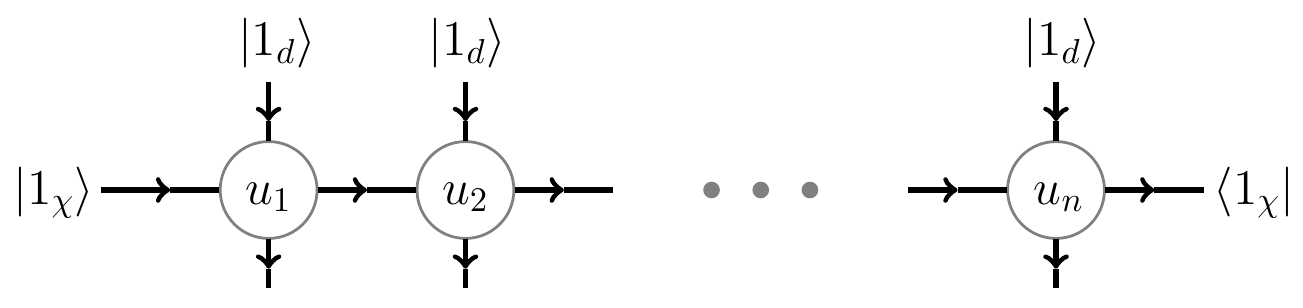} 
\caption{Possible preparation of an $\textsf{MPS}$ ($\textsf{OBC}$). The $n$-particle chain and an ancilla are all prepared in some product reference state, sequential unitary interaction next occurs, the ancilla is eventually projected and disregarded.}\label{fig:seq-int}
\end{center}
\end{figure}

\emph{\textbf{The Haar-Markov chain.--}} Let us consider a system of $n$ identical 'spin' particles confined on a line segment, interacting via some nearest-neighbour Hamiltonian $H$. From now on, we will assume open boundary conditions ($\textsf{OBC}$) and we will denote $d$ the dimension of the Hilbert space $\mathcal{H}$ describing each particle. We aim at approximating the ground state of $H$ with a matrix product state ($\textsf{MPS}$), i.e. a state of the form
\beq\label{eq:def-mps}
\ket{\psi}=\sum_{s_1 \ldots s_n} \tr[A_1(s_1) \ldots A_{n}(s_n)] \ket{s_1 \ldots s_n},
\eeq
where the matrices $A_1(s_1)$ have size $1 \times \chi$, the matrices $A_k(s_k), \; k=2 \ldots n-1$ have size $\chi \times \chi$, and the matrices $A_n(s_n)$ have size $\chi \times 1$. The integer $\chi$, called bond dimension, is a parameter that governs the amount of entanglement that can be supported by the state. $s_k \in \{1, \ldots, d\}$ is a local degree of freedom associated with particle $k$. A possible way to prepare states of the form (\ref{eq:def-mps}) is through unitary sequential interaction between an ancilla and an array of $n$ particles at fixed locations, as indicated on Fig.\ref{fig:seq-int}. The fact that \emph{any} $\textsf{MPS}$ can be prepared in this way \cite{Solano}  allows to define Markov chains on the configuration space $\Omega \equiv U(d \chi)^{\otimes n}$ without loss of generality; each tensor $A_k$ can be viewed as a piece of some $d\chi \times d\chi$ unitary matrix $u_k$. The (unique) quantum state associated with a configuration $\omega=\{u_k: k=1 \ldots n\}$ will be denoted $\ket{\omega}$.

\begin{figure}[h]
\begin{center}
\includegraphics[width=70mm,height=15mm]{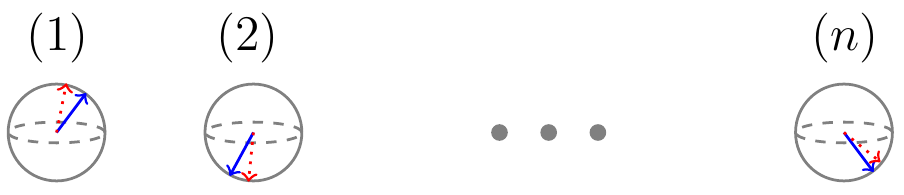} 
\caption{(Color online.) Schematic illustration of a Haar-Markov move for an $n$-particle chain. In this example, where $d=2$ and $\chi=1$, the configuration space is $U(2)^{\otimes n}$ and can be represented as a set of $n$ points on $2-$spheres. The set of plain (resp. dotted) arrows represents the current (resp. candidate) configuration.}\label{fig:illu-MH}
\end{center}
\end{figure}

Let $h(\omega)=\bra{\omega} H \ket{\omega}/\braket{\omega}{\omega}$ denote the energy associated with a configuration $\omega$. We will be interested in Boltzmann probability distributions of the form
\beq\label{eq:Boltzmann}
\pi^{(\beta)}(\omega)\equiv \frac{e^{-\beta h(\omega)}}{\int d\omega' e^{-\beta h(\omega')}}, \; \omega \in \Omega,
\eeq
where $d\omega$ denotes the Haar measure over $U(d \chi)^{\otimes n}$. The ability to efficiently draw samples according to $\pi^{(\beta)}$ for large values of $\beta$ would be an invaluable resource and would immediately lead to huge progress on the optimal state problem ; as $\beta$ increases, the support of $\pi^{(\beta)}$ exponentially concentrates on low energy configurations. However, except for a few trivial cases, \emph{direct} sampling of $\pi^{(\beta)}$ seems to be impossible. It is then natural to construct a Markov chain that produces samples (approximately) distributed according to $\pi^{(\beta)}$. A simple such Markov chain is defined by the following transition rule (\emph{Haar-Markov}): 

\begin{itemize}

\item {Draw $n$ i.i.d. unitary matrices $\{v_k, k=1 \ldots n \} \in U(d\chi)^{\otimes n}$ according to the Haar measure \cite{mezza}.} 

\item {From the current configuration $\omega=\{u_k, k=1 \ldots n\}$, construct a can\-di\-da\-te $\omega'=\{u_k v_k^{\alpha}: k=1  \ldots n\}$, $0 < \alpha <1.$ (See Illustration on Fig. \ref{fig:illu-MH}.)} 

\item {Accept the proposed move $\omega \to \omega'$ with Metropolis probability:}
\beq\label{eq:Met}
P^{(\beta)}_{\text{acc}}(\omega \to \omega')=\min \{ 1, e^{-\beta (h(\omega')-h(\omega))}\}. 
\eeq

\end{itemize}

This Markov chain is strongly ergodic and satisfies the celebrated detailed balance condition. It  is furthermore defined on a compact configuration space; these conditions suffice to establish convergence (see Appendix \ref{app:conv} for a brief discussion). We also note that the proposed moves are global; they involve \emph{each} particle of the system. 

When $\beta$ is varied with time $t$, in such a way that $\lim_{t \to \infty} \beta(t)=\infty$, the concatenated Markov chains result in a simulating annealing scheme. Choosing the value of the parameter $\alpha$ is a delicate matter \cite{roberts}; an appropriate tradeoff has to be found between ambitious and modest moves. We have observed that the Haar-Markov chain is indeed sensitive to this parameter; good performances were obtained when $\alpha$ is tuned so that the average acceptance rate of the proposed moves is about $1/4$. The Haar-Markov chain is, by construction, sign problem free for \emph{any} Hamiltonian, even for fermionic systems (see below). We believe this property is important. Of course, we expect the complexity of some difficult Hamiltonians will manifest itself in the form of long convergence times, but there is no ill-defined probabilities here.

\emph{\textbf{A case study: the Ising model.--}} We have tested the Markov chain described in the previous section on the following Ising Hamiltonian:
\beq\label{eq:Ising-H}
H=-\sum_{k=1}^{n-1} \sigma^x_k \sigma^x_{k+1}-\sum_{k=1}^{n} \sigma^z_k.
\eeq

It is critical in the thermodynamic limit. We have considered a system made of $n=40$ particles, and have aimed at estimating its ground state energy using $\textsf{MPS}$ with bond dimension equal to $2,4,8,16$. A crucial issue is to decide the rate at which $\beta$ should be increased from one step to the next. Motivated by early results on fast simulated annealing \cite{FSA}, we have been interested in testing the power of linear schedules. The initial inverse temperature was set to $\beta_0=1$ and increased at each iteration by an equal amount $d\beta=0.05$. We make no claim of optimality regarding this schedule; possibly more ambitious choices, such as an exponential schedule for instance, could lead to faster algorithms. Our results are summarised on Table \ref{table:results-Ising-1D}. The estimates for the ground state energy are comparable to those output by Euclidean evolution. A remark is in order regarding how the figures should be interpreted; we have not aimed at getting results exact up to, say, ten digits. Rather, just like other simulated annealing schemes used in other contexts, our scheme should be viewed as a tool to get \emph{close enough} to a \emph{global} optimum, that when completed with a more greedy method, it yields the correct solution. 

\begin{table}[here]
\begin{tabular}{cccc}
$\chi$ & $\mathsf{HM} \; (\beta < \infty)$ & $\mathsf{HM} \; (\beta=\infty)$ & $\mathsf{EE}$ \\
\hline
\hline
2 & $-50.354476$ &  $-50.503292$   & $-50.503303$\\
4 & $-50.489549$ &  $-50.511711$   & $-50.567694$\\
8 & $-50.338708$  & $-50.514589$ &   $-50.569413$\\
16 & $-50.354480$ & $-50.502324$ &  $-50.569425$\\
\end{tabular}
\caption{Ground state energies for the Ising model using Haar-Markov chains. The first column records the bond dimension used in the computations. The second column reports the estimate for the ground state energy output by our simulated annealing scheme. The state obtained after simulated annealing is then used as a seed for Markov chains at zero temperature. The resulting ground state energies are shown on the third column. The fourth column shows the results obtained through Euclidean evolution for the sake of comparison. 
}
\label{table:results-Ising-1D}
\end{table}

\emph{\textbf{Extension to two dimensions.--}} The Haar-Markov chains defined above carry through to two-dimensional systems. Let us consider an $n_x \times n_y$ square lattice ($\textsf{OBC}$ assumed). First, a unitary configuration space $\Omega$ can again be associated with the set of all $\textsf{PEPS}$ with a fixed bond dimension $\chi$. One simply observe that each row of a $\textsf{PEPS}$ $\psi$ can be viewed as an $\textsf{MPS}$ with "physical dimension" $d'=d\chi^2$, after reshuffling the indices of the tensors defining $\psi$. As a result, the choice of configuration space introduced in one dimension applies to each \emph{row} of a $\textsf{PEPS}$, and shows there is an onto map from $U(d \chi^3)^{\otimes n_x n_y}$ to the set of $\textsf{OBC}$ $\textsf{PEPS}$. The next issue is to define a cost function. The energy of a configuration $\omega$ can generally not be evaluated exactly. Therefore, we should not aim at approximating the energy itself, but some suitable approximation. There is some freedom when making a choice for this replacement. The details and motivations for our own choice, which we will denote $\check{h}(\omega)$, are given in Appendix \ref{app:two-dim-too}. Importantly, $\check{h}$ is by construction a \emph{rigorous} upper bound on the ground state energy of $H$ for \emph{any} Hamiltonian. As a result, the instability issue studied in Ref.\cite{unstable-peps} simply never shows up. We again consider distributions of the form given by Eq.(\ref{eq:Boltzmann}), with $h$ repaced with $\check{h}$; strong ergodicity and reversibility are still satisfied. Thus, for a sufficiently slow cooling schedule, the algorithm \emph{must} again converge to a global minimum of $\check{h}$ over $\Omega$. We have tested our scheme in the case of the Ising model 
\beq
H=-J_x \sum_{\langle k,l \rangle} \sigma^x_k \sigma^x_l- h_z \sum_k \sigma^z_k, 
\eeq
for which our computations could be compared to a Euclidean evolution \cite{Lub}. Our results displayed on Table \ref{table:results-Ising-2D} are very encouraging. An advantage of the present scheme, when compared to other search methods, is its relative simplicity; it completely skips sophisticated computations of single-site environments \cite{peps}.

\begin{table}[here]
\begin{tabular}{cccc}
$J_x$ & $h_z$ & $\textsf{HM}$ & $\textsf{EE}$\\
\hline
\hline
$1$ & $1/2$  & $-85.825$ & $-85.869(3)$\\
$1$ & $1$ & $-91.492$ & $-91.520(8)$\\
$1/2$ & $1$  & $-57.192$ & $ -57.377(3)$\\
\end{tabular}
\caption{$\textsf{PEPS}$ computations for the 2D Ising model ground state energy. A square $7 \times 7$ lattice was considered. The third column relates to our computations, the 4th to Euclidean evolution. Bond dimension equal to $2$; rank $\chi_c$ of matrix product boundary operators equal to $4$. 
}\label{table:results-Ising-2D}
\end{table}

\emph{\textbf{Discussion.--}} In summary, we have introduced a stochastic algorithm to search optimal tensor network states, and demonstrated its validity in one and in two dimensions. This algorithm is rooted in strongly ergodic reversible Markov chains defined on a connected compact configuration space. Therefore, one can prove that convergence to a global optimum is always possible for \emph{some} sufficiently slow schedule. 
A deeper problem would be to bound the \emph{mixing time} \cite{mixing} of the Haar-Markov chains at fixed temperature, even in one dimension and for simple Hamiltonians. 

The results we have obtained are encouraging, and we believe the class of Markov chains introduced here deserve further study. Our method has turned out to be slower than, say Euclidean evolution, but not prohibitively. Moreover, the algorithm could be improved in many ways; faster schedules and finer results should be feasible. For instance, departing from the Haar measure to pick a candidate might boost the global convergence properties; in particular, biased priors in spirit of the Metropolis-Hastings algorithm are worth considering (see below). Other choices of configuration space might also improve the global performances. An important clarification should however be made regarding the purpose we should pursue. We have not sought to outperform the greedy techniques mentioned in the introduction; these are as fast as can be and provide excellent results in their range of validity. Rather, we believe our heuristics should ideally be used as a \emph{preconditioner}, supplying a good 'seed' to greedy methods, when necessary. An advantage of our scheme is its great flexibility with respect to the objective function; for example, while existing methods are mainly aimed at investigating low energy properties of hamiltonians, the present method could in principle be used to investigate high energy states for instance (see below). Another aspect of the simulated annealing scheme is its ease of implementation. This is particularly true in two dimensions, where the avoidance of single site environment computations is a significant simplification. 

We conclude with a brief description of ideas stemming from the present work that will be analysed elsewhere.

\emph{{Periodic boundary conditions.-}} Just as we saw in the case of $\textsf{PEPS}$, a simple reshuffling of indices allows to regard any periodic boundary condition ($\textsf{PBC}$) $\textsf{MPS}$ \cite{review-mps} as an $\textsf{OBC}$ $\textsf{MPS}$ with effective physical indices of higher rank $d'=d\chi$ for two 'edge' sites. The same algorithm could therefore be used to study systems with boundaries identified.

\emph{{Low energy eigenstates.-} }Let $\{ \ket{\phi_0}, \ldots, \ket{\phi_{l-1}}$ denote $\textsf{MPS}$ approximations for the first $l$ eigenstates of some Hamiltonian $H$. An approximation scheme for the $(l+1)$-th eigenstate can be simply constructed from the modified figure of merit
\bed
h_{\text{LS}}(\omega)=h(\omega)+\lambda \sum_{k < l} |\braket{\phi_k}{\omega}|,
\eed
where $\lambda >0$ is a Lagrange multiplier, which value can be chosen at our convenience.  

\emph{{High energy eigenstates.-}} Assume we wish to find an approximation to \emph{some} eigenstate with energy in an interval 
$[e_1,e_2]$. A possibility could be to minimise the following objective function:
\bed
h_{\text{HS}}(\omega)=\theta(h(\omega)-e_1) \theta(e_2-h(\omega))+\lambda [\bra{\omega} H^2 \ket{\omega}-\bra{\omega} H \ket{\omega}^2],
\eed
where $\theta$ denotes the Heaviside step function, $\lambda >0$ denotes again some Lagrange multiplier.

\emph{{Spectral decomposition and time evolution.-}} Consider some hamiltonian $H$, some $\textsf{PEPS}$ $\ket{\phi}$, and let 
\bed
\ket{\phi}=\sum_i c_i \ket{\psi_i}
\eed
denote the spectral decomposition of $\ket{\phi}$ in terms of the eigenstates of $H$. To construct $\textsf{PEPS}$ approximations for the states appearing in this decomposition, one could try to optimise the figure of merit
\bed
h_{\text{SD}}(\omega)=|\braket{\omega}{\phi}|+\lambda [\bra{\omega} H^2 \ket{\omega}-\bra{\omega} H \ket{\omega}^2], \; \lambda <0.
\eed
When a solution $\omega^{\star}_1$ is found, one repeats the optimisation with the state 
\bed
\ket{\phi}-\braket{\omega^{\star}_1}{\phi} \ket{\omega^{\star}_1}.
\eed 
And so on. Proceeding in this way, \`a la Gram-Schmidt, one sequentially constructs an approximate spectral decomposition: 
\bed
\ket{\phi'}=\sum_{i} \braket{\omega^{\star}_i}{\phi} \ket{\omega^{\star}_i}
\eed
for $\ket{\phi}$. Time evolution for $\phi$ is next approximated as
\bed
e^{-iH t} \ket{\phi}=\sum_{i} \braket{\omega^{\star}_i}{\phi} \; e^{-i \bra{\omega^{\star}_i} H\ket{\omega^{\star}_i} t } \; \ket{\omega^{\star}_i}.
\eed

\emph{{Two-dimensional Hamiltonians at finite temperature.-}} Let $\{e_i, \ket{\psi_i} \}$ denote the spectral decomposition of some two-dimensional quantum Hamiltonian, and assume we are interested in the mean value of some observable $A$ at finite temperature:
\bed
\mean{A}=\frac{1}{Z} \sum_i e^{-e_i/T} \bra{\psi_i} A \ket{\psi_i},
\eed
 where $Z=\sum_i e^{-e_i/T}$ denotes the partition function of the system. Let $\alpha_1, \ldots, \alpha_M$  denote samples representative of the distribution $p_T(i)= e^{-e_i/T}/Z$. $\mean{A}$ can be estimated as 
 \bed
 \mean{A} \simeq \frac{1}{M} \sum_{j=1}^M \bra{\psi_{\alpha_j}} A \ket{\psi_{\alpha_j}}.
 \eed
The law of large numbers ensures the average error made with this estimate should decrease as $1/{\sqrt{M}}$. Yet another Markov chain can be designed to obtain approximations to such representative states. Let $\kappa$ denote an upper bound on $||H||_{\infty}$. Such an upper bound can generally be constructed easily. Let $\phi$ denote the current ($\mathsf{PEPS}$) state of the Markov chain, and let $\check{h}(\phi)$ denote an approximation for $\bra{\phi} H \ket{\phi}/\braket{\phi}{\phi}$.
\begin{itemize}

\item Draw a small positive random number $\delta e$. 

\item Use the method described above to try and construct an approximate eigenstate $\phi'$ of $H$ with approximate energy $\check{h}(\phi')$ in $[\check{h}(\phi)-\delta e, \check{h}(\phi)+\delta e] \cap [-\kappa,+\kappa]$.

\item If such a state $\phi'$ can be constructed, the state of the Markov chain is updated to $\phi'$ with Metropolis probability $\text{min} \{1, e^{-(\check{h}(\phi')-\check{h}(\phi))/T}\}$.

\end{itemize}

\emph{{Entanglement Renormalisation.-}} We have here limited ourselves to $\textsf{MPS}$ and $\textsf{PEPS}$, but the analysis evidently carries through to self-similar tensor network states \cite{mera-cft}. Configuration spaces of the form we have used are actually very natural in that context, because it is demanded that the tensors defining the state be unitary.   

\emph{{Fermions.-}} The notions of $\textsf{MPS}$ and $\textsf{PEPS}$ have been extended to the case of fermionic degrees of freedom \cite{Fermionic-PEPS}. In particular, the mean value of any operator for a system in a pure fermionic $\textsf{PEPS}$ can be (efficiently) translated into the mean value of a $\textsf{PEPS}$ defined on an associated bosonic system. Our Markov chains therefore carry through to the fermionic case. Since the mean of a (fermionic) Hamiltonian will be by construction a \emph{real} number, there is still no sign problem.

\emph{{Biased priors.-}} Conceivably, the convergence of the Markov chain could be significantly boosted if the candidate were not drawn 'isotropically', but according to some well chosen biased prior $\pi^{(\beta)}_{\textrm{pick}}(\omega'|\omega)$. Reversibility could still be satisfied if the Metropolis-Hastings rule were used to decide acceptance \cite{MH}:
\beq
P_{\textrm{acc}}^{(\beta)}=\min \{1, \frac{\pi^{(\beta)}(\omega')}{\pi^{(\beta)}(\omega)} \times 
\frac{\pi^{(\beta)}_{\textrm{pick}}(\omega|\omega')}{\pi^{(\beta)}_{\textrm{pick}}(\omega'|\omega)} \}.
\eeq 

The construction of such biased priors is currently under investigation.

\emph{\textbf{Acknowledgements.--}} I am grateful to Jos\'e Ignacio Latorre, David P\'erez-Garc\'ia, Mauricio Mari\~no, Mari-Carmen Ba\~nuls, Roman Or\'us, Angelo Luc\'ia, Germ\'an Sierra, Luca Tagliacozzo and Andrew Ferris for discussions and useful comments at various stages of the present work. I also thank Michael Lubasch for having provided me with energy estimates that made the comparisons of  
Table \ref{table:results-Ising-2D} possible. Financial support from the Ram\'on y Cajal programme (RYC-2009-04318), Spanish Grant FIS2010-16185, and Grup de Recerca Consolidat (Generalitat de Catalunya) 2009 SGR21, is acknowledged.



\appendix

\section{Convergence}\label{app:conv}

The Haar-Markov chain is strongly ergodic; regarding $m=O(1/\alpha)$ consecutive steps of one such chain as a single super-step, one easily sees that the probability for a transition between any pair of configurations $(\omega,\omega')$ is strictly positive. Also, the acceptance rule satisfies the celebrated detailed balance condition (reversibility)
\beq\label{eq:db}
\pi^{(\beta)}(\omega) P^{(\beta)}_{\text{acc}}(\omega \to \omega')=\pi^{(\beta)}(\omega') P^{(\beta)}_{\text{acc}}(\omega' \to \omega).
\eeq
The configuration space is moreover compact. (It is also connected.) The general theory related to the Metropolis algorithm \cite{Bhanot} can therefore be invoked to claim that, for any fixed $\beta$, any initial probability distribution $\pi_0$ is mapped after $\tau$ steps to a distribution $\pi_{\tau}$ which is exponentially close (in total variation distance) to the target distribution: 
\bed
||\pi_{\tau}-\pi^{(\beta)}||_{\textrm{TV}} \leq 2 \; \eta(\beta)^{\tau}.
\eed
The constant $\eta(\beta)$ is guaranteed to be strictly lower than $1$. Thus  $||\pi_{\tau}-\pi^{(\beta)}||_{\textrm{TV}}$ can be made lower than any constant $\delta > 0$ in $O(\ln(1/\delta))$ steps. Finding a useful upper bound for $\eta(\beta)$ is however a difficult problem in general \cite{mixing}. As stated in the introduction, it has recently been proven that for one-dimensional gapped Hamiltonians and $\textsf{MPS}$, the optimal state problem has polynomial complexity. In the present context, bounds on $\eta(\beta)$ combined with simulated annealing could lead to an alternative proof of that result, together with a very simple approximation scheme. 

\section{Extension to two dimensions: addendum}\label{app:two-dim-too}

We start with some general considerations regarding $\textsf{PEPS}$ for an $n_x \times n_y$ square lattice system with open boundary conditions. The mean value of any observable $Y$ for a $\textsf{PEPS}$ $\psi$ can be expressed as 
\beq\label{eq:peps-mean-value}
\bra{\psi} Y \ket{\psi}=\bra{\partial_{\text{top}}} \prod_{k=1}^{n_y-2} \mathcal{T}_k  \ket{\partial_{\text{bot}}}. 
\eeq
where the boundary 'states' $\ket{\partial_{\text{top}}},\ket{\partial_{\text{bot}}}$ are $\textsf{MPS}$, and the transfer operators $\mathcal{T}_k$ are matrix product operators \cite{peps}, each with a fixed bond dimension. The r.h.s. can generally not be evaluated exactly in a time that grows polynomially with $n_x,n_y$ and the bond dimension. It is thus common to evaluate the r.h.s. of Eq.(\ref{eq:peps-mean-value}) approximately as follows.
\begin{enumerate}

\item A 'bottom' edge $\textsf{MPS}$ $\ket{\partial'_{\text{bot}}(0)}=\ket{\partial_{\text{bot}}}$ is initialised.

\item $\forall k=1 \ldots n_y-2$, an $\textsf{MPS}$ $\ket{\partial'_{\text{bot}}(k)}$, with fixed bond dimension $\chi_c$, approximating $\mathcal{T}_{n_y-k-1} \ket{\partial'_{\text{bot}}(k-1)}$, is constructed.

\item The amplitude $\braket{\partial_{\text{top}}}{\partial'_{\text{bot}}(n_y-2)}$ is eventually proposed as an estimate for $\bra{\psi} Y \ket{\psi}$.

\end{enumerate}

We have not proceeded differently. The following lemma, which is an easy corollary of the triangular and the Cauchy-Schwarz inequalities, provides a bound on the error resulting from this approximate contraction scheme. 
\begin{lemma} Let $\mathbb{V}$ denote a finite-dimensional vector space. For any finite sequence of elements $\{v_1, \ldots v_m\}  \subset \mathbb{V}$,
\beq\label{eq:lemma-peps-1}
||v_1-v_m|| \leq \sum_{j=1}^{m-1} ||v_j-v_{j+1}||. 
\eeq
Moreover, $\forall u,v,u',v' \in \mathbb{V}$,
\beq\label{eq:lemma-peps-2}
|(u',v')-(u,v)| \leq ||u'|| \cdot ||v-v'||+||u-u'|| \cdot (||v'||+||v-v'||).
\eeq
\end{lemma}

The inequality (\ref{eq:lemma-peps-1}) allows to sequentially compute an estimate for how much $\ket{\partial'_{\text{bot}}(n_y-2)}$ and $ \prod_{k=1}^{n_y-2} \mathcal{T}_k  \ket{\partial_{\text{bot}}}$ differ, while the inequality (\ref{eq:lemma-peps-2}) allows to use this estimate to bound the error on $\bra{\partial_{\text{top}}} \prod_{k=1}^{n_y-2} \mathcal{T}_k  \ket{\partial_{\text{bot}}}$. These considerations are relevant to our Monte-Carlo scheme, when applied to $\textsf{PEPS}$. When dealing with a concrete Hamiltonian $H$ and a configuration $\omega \in \Omega$, they provide an estimate $h'(\omega)$ for $\bra{\omega} H \ket{\omega}$, as well as an upper bound $\delta_h(\omega)$ on the error. Similarly, an estimate $n'(\omega)$ for the square norm $\braket{\omega}{\omega}$, together with an upper bound $\delta_n(\omega)$ on the norm error, can be computed. The presence of these errors $\delta_h(\omega)$ and $\delta_n(\omega)$ is a major distinction between $\textsf{MPS}$ and $\textsf{PEPS}$; the following definition is introduced to guarantee they are properly taken into account. 

\begin{definition} 
A $\textsf{PEPS}$ configuration $\omega$ is \textbf{pointless} at $\chi_c$ if $\delta_h(\omega) \geq h'(\omega)$,  $\delta_n(\omega) \geq n'(\omega)$, or both.
\end{definition}
We are now in a position to write down the objective function we have used in our concrete computations:
\beq\label{eq:PEPS-FOM}
\check{h}(\omega)= \left\{
\begin{array}{rl} 
&+\infty  \; \text{if $\omega$ is pointless} ,\\ 
&\max_{a,b \in \{0,1\}} \{ \frac{h'(\omega)+(-)^{a} \delta_h(\omega)}{n'(\omega)+(-)^{b} \delta_n(\omega)} \} \; \text{else}.
\end{array} \right. 
\eeq
The rationale behind this choice is most easily explained with a comparison between two extreme cases. Given a Hamiltonian $H$, and having to choose between an \emph{exact} description of its ground state in terms of some $\textsf{PEPS}$ $\psi_0$, which cannot be contracted reliably, and an \emph{approximate} description $\ket{\phi_0}$ we can work with, the latter option should be preferred. 

Plugging $\check{h}(\omega)$ into the Boltzmann weights (\ref{eq:Boltzmann}) guarantees that the random approximation scheme will provide \emph{rigorous} upper bounds on the ground state energy of $H$. The use of $\check{h}(\omega)$ effectively steers Markov-chains away from pointless $\textsf{PEPS}$. As a result, none of the instability issues addressed in Ref.\cite{unstable-peps} has to be faced.

\end{document}